\documentclass[prd,amsmath,amssymb,showpacs,superscriptaddress,nofootinbib,twocolumn]{revtex4-1}
\usepackage{amsmath}
\usepackage{amssymb}
\usepackage{graphicx}
\usepackage{dcolumn}
\usepackage{bm}
\usepackage{color}
\usepackage{subfigure}
\usepackage{xspace}
\usepackage{textpos}
\usepackage[english]{babel}
\usepackage[utf8]{inputenc}
\usepackage{amsmath,linnum}
\usepackage[mathlines]{lineno}
\usepackage{color}
\usepackage[bookmarksnumbered, pdfstartview=FitH,colorlinks,linkcolor=blue,anchorcolor=blue,citecolor=blue]{hyperref}
\usepackage[perpage,symbol]{footmisc}
\usepackage{subfigure}
\usepackage{indentfirst}
\usepackage{fancyvrb}
\usepackage{multirow}
\usepackage{float}
\setfnsymbol{wiley}
\usepackage{epstopdf}

\renewcommand{\thefootnote}{\fnsymbol{footnote}}

\begin{document}

\title{\bf \boldmath Dalitz Plot Analysis of the Decay $\omega \rightarrow \pi^{+}\pi^{-}\pi^{0}$}

\author{
  \begin{small}
    \begin{center}
      M.~Ablikim$^{1}$, M.~N.~Achasov$^{10,d}$, S. ~Ahmed$^{15}$, M.~Albrecht$^{4}$, M.~Alekseev$^{55A,55C}$, A.~Amoroso$^{55A,55C}$, F.~F.~An$^{1}$, Q.~An$^{52,42}$, Y.~Bai$^{41}$, O.~Bakina$^{27}$
% ...
      \\
      \vspace{0.2cm}
      (BESIII Collaboration)\\
      \vspace{0.2cm} {\it
$^{1}$ Institute of High Energy Physics, Beijing 100049, People's Republic of China\\
% ...
$^{4}$ Bochum Ruhr-University, D-44780 Bochum, Germany\\
% ...
$^{10}$ G.I. Budker Institute of Nuclear Physics SB RAS (BINP), Novosibirsk 630090, Russia\\
% ...
$^{15}$ Helmholtz Institute Mainz, Johann-Joachim-Becher-Weg 45, D-55099 Mainz, Germany\\
% ...
$^{27}$ Joint Institute for Nuclear Research, 141980 Dubna, Moscow region, Russia\\
% ...
$^{41}$ Southeast University, Nanjing 211100, People's Republic of China\\
$^{42}$ State Key Laboratory of Particle Detection and Electronics, Beijing 100049, Hefei 230026, People's Republic of China\\
% ...
$^{52}$ University of Science and Technology of China, Hefei 230026, People's Republic of China\\
% ...
$^{55}$ (A)University of Turin, I-10125, Turin, Italy; (B)University of Eastern Piedmont, I-15121, Alessandria, Italy; (C)INFN, I-10125, Turin, Italy\\
% ...
$^{d}$ Also at the Novosibirsk State University, Novosibirsk, 630090, Russia\\
% ...
      }
    \end{center}
    \vspace{0.4cm}
  \end{small}
}
\noaffiliation{}

\begin{abstract}
  Using a low-background sample of $2.6\times 10^5$ $J/\psi\rightarrow\omega\eta(\omega\rightarrow\pi^{+}\pi^{-}\pi^{0},\eta\rightarrow\gamma\gamma)$ events, about 5 times larger statistics than previous experiments, we present a Dalitz plot analysis of the decay $\omega\rightarrow\pi^{+}\pi^{-}\pi^{0}$.  It is found that the Dalitz plot distribution differs from the pure $P$-wave phase space with a statistical significance of $18.9\sigma$. The parameters from the fit to data are in reasonable agreement with those without the cross-channel effect within the dispersive framework, which indicates that the cross-channel effect in $\omega\rightarrow\pi^+\pi^-\pi^0$ is not significant.
\end{abstract}


\maketitle

\section{Introduction}\label{sec:introduction}

At low energies, the process $e^+e^-\to\pi^+\pi^-\pi^0$ is dominated by the contributions from the $\omega$ or $\phi$ isoscalar vector mesons.
The precise knowledge of the reaction is needed for the determination of the hadronic contribution to the muon anomalous magnetic moment $(g-2)_\mu$ \cite{Jegerlehner:2009ry}.
The  $e^+e^-\to\pi^+\pi^-\pi^0$ process provides the second-most important contribution to the hadronic vacuum polarization.
In addition, the differential distribution of the pions is an important benchmark for the determination of the dominant part of the hadronic light-by-light contribution to $(g-2)_\mu$ originating from the $\pi^0$ meson pole using dispersion theory~\cite{Hoferichter:2014vra}.
Since the study of the Dalitz plots of $\omega/\phi\rightarrow\pi^+\pi^-\pi^0$ can provide further constraints to the calculation of the electromagnetic transition form factors of
$\omega/\phi\rightarrow\pi^0\gamma^{\star}$~\cite{Schneider2012,Danilkin:2014cra},
both $\omega\rightarrow\pi^+\pi^-\pi^0$ and $\phi\rightarrow\pi^+\pi^-\pi^0$ still attract attention of both theorists and experimentalists.
Within the dispersive framework~\cite{Niecknig:2012sj,Danilkin:2014cra}, the Dalitz plot distributions of these two decays and integrated decay widths are presented.
It was found that the dispersive analysis can provide a good description of the precise $\phi\rightarrow\pi^+\pi^-\pi^0$ data from KLOE experiment~\cite{Aloisio:2003ur}.
However, no experimental data of comparable precision on $\omega\rightarrow\pi^+\pi^-\pi^0$ exists to compare with the predictions.
The $\omega\rightarrow\pi^+\pi^-\pi^0$ decay can be described in the Isobar Model as proceeding via an intermediate $\rho\pi$ state.
In addition, the third pion can interact with the decay products of the $\rho$ resonance.
This so-called crossed-channel effect~\cite{Danilkin:2014cra} is predicted to provide a significant contribution to the decay and should modify the Dalitz plot distribution.
The recent Dalitz plot analysis from WASA-at-COSY Collaboration of $\omega \rightarrow \pi^{+}\pi^{-}\pi^{0}$~\cite{Adlarson:2016wkw} with a combined sample of $4.4\times 10^4$ events has given evidence of final-state interaction (FSI) in this channel.

The $\omega$ meson is abundantly produced in $J/\psi$ decays, with an overall branching fraction of $1\%$. The world's largest sample of $1.3\times 10^9$ $J/\psi$ events collected with the BESIII detector offers a unique opportunity to investigate the Dalitz plot of $\omega\rightarrow \pi^+\pi^-\pi^0$. In this paper, the two-body decay $J/\psi\rightarrow\omega\eta$ is used to select a clean sample of $\omega$ events. This two-body decay not only has a large branching fraction of $(1.74\pm0.20)\times 10^{-3}$~\cite{Patrignani:2016xqp}, but also provides a very simple event topology, in which the $\omega$
can be tagged by the $\eta$ meson dominant decay mode into two photons.

In this analysis, we construct the Dalitz plot of $\omega \rightarrow \pi^{+}\pi^{-}\pi^{0}$ using the dimensionless variables defined in Ref.~\cite{Schneider:2010hs},
\begin{equation}
x=\frac{t-u}{\sqrt{3}R_{\omega}},\ \ \ \ y=\frac{s-s_{0}}{R_{\omega}}+\frac{2(m_{\pi^{\pm}}-m_{\pi^{0}})}{m_{\omega}-2m_{\pi^{\pm}}-m_{\pi^{0}}},\label{eq:5}
\end{equation}
where $s,\ t,\ u$ are the invariant masses squared of the $\pi^{+}\pi^{-}$, $\pi^{-}\pi^{0}$, and $\pi^{0}\pi^{+}$ systems, respectively; $s_{0}=(s+t+u)/3$, $R_{\omega}=\frac{2}{3}m_{\omega}(m_{\omega}-m_{\pi^{+}}-m_{\pi^{-}}-m_{\pi^{0}})$.
Alternatively, for a description of an isospin conserving process the related polar variables $z$ and $\phi$ can be used,
\begin{equation}
z=\left |x+yi\right |^2,\ \phi=\arg(x+yi).\label{eq:6}
\end{equation}

In accordance with Ref.~\cite{Niecknig:2012sj}, the density of the Dalitz plot for $\omega \rightarrow \pi^{+}\pi^{-}\pi^{0}$ can be written as
\begin{equation}
\left | {\cal{M}}\right |^{2}=\frac{\left | \vec{p}_{+}\times \vec{p}_{-}\right |^{2}}{m_{\omega}}\cdot\left | {\cal{F}}\right |^{2},\label{eq:2}
\end{equation}
where $\vec{p}_{+}$ and $\vec{p}_{-}$ are the momenta of $\pi^{+}$ and $\pi^{-}$ in the $\omega$ rest frame, respectively.
If there is no FSI in this decay, ${\cal{M}}$ is distributed like $P$-wave phase space, with $|{\cal{F}}|^2=1$.
But for $\omega \rightarrow \pi^{+}\pi^{-}\pi^{0}$, ${\cal{F}}$ can be described by the Omn\`es function,
which is a function of $s$, $t$ and $u$ and is calculated by a dispersive analysis~\cite{Niecknig:2012sj}.
Since $s,t,u$ can be transformed into $z$ and $\phi$, this function can also be asymptotically expanded into a polynomial of the variables $z$ and $\phi$:
\begin{equation}
\begin{split}
\left|{\cal{F}}(z,\phi)\right|^{2}\propto&
1+2\alpha z+2\beta z^{3/2}\sin3\phi \\
&+2\zeta z^{2}+2\delta z^{5/2}\sin3\phi+{\cal{O}}(z^{3}),\label{eq:3}
\end{split}
\end{equation}
where $\alpha,\beta,\zeta,\delta$ are parameters to be determined by a fit to data.
This parametrization conserves isospin in the amplitude, which is equivalent to invariance under the transformation $\phi\to\phi+120^\circ$.

\section{Detector and MONTE CARLO SIMULATION}\label{sec:detector}

BEPCII is a double-ring $e^{+}e^{-}$ collider working at center-of-mass energies from 2.0 to 4.6 GeV. The BESIII detector~\cite{Ablikim2010}, with a geometrical acceptance of 93\% of 4$\pi$ stereo angle, operates in a 1.0~T (0.9~T in 2012, when about 83\% of the data sample were collected) magnetic field provided by a superconducting solenoid magnet. The detector is composed of a helium-based drift chamber (MDC), a plastic scintillator time-of-flight (TOF) system, a CsI(Tl) electromagnetic calorimeter (EMC) and a muon counter (MUC) consisting of resistive plate chambers (RPC) interleaved in the steel of the flux return yoke. The charged-particle momentum resolution at 1.0 GeV/$c$ is 0.5\%, and the specific energy loss ($dE/dx$) resolution is better than 6\%. The time resolution of the TOF is 80~ps in the barrel and 110~ps in the endcaps. The energy resolution of the EMC at 1.0~GeV/$c$ is 2.5\% (5\%) for electrons and photons in the barrel (endcaps), and the position resolution is better than 6 mm (9 mm) in the barrel (endcaps). The position resolution in the MUC is better than 2 cm.

Monte Carlo (MC) simulated event samples are used to estimate backgrounds and
determine the detection efficiencies. The GEANT4-based
\cite{Agostinelli2003} simulation software BOOST \cite{Deng2006}
includes the geometric and material description of the BESIII detector,
the detector response, and digitization models, as well as the information on the running conditions and the detector
performance.
In this analysis, three-body decays without FSI are generated by a phase space generator (PHSP) in which events are produced with uniform distribution in their Dalitz plot.
The production of the $J/\psi$ resonance is simulated
with the MC generator {\sc kkmc}~\cite{Jadach2000,Jadach2001}, while the decays are
generated with EVTGEN \cite{Lange2001} with branching fractions being
set to the world average values \cite{Patrignani:2016xqp} for the
known decay modes, and with LUNDCHARM \cite{Chen2000,Ping2008} for the remaining
unknown decays. We use an inclusive sample of $1.2 \times 10^{9}$
simulated $J/\psi$ events to identify background contributions.

\section{Event Selection}\label{sec:selection}

The $J/\psi\rightarrow\omega\eta,\ \omega\to\pi^+\pi^-\pi^0$ events
are reconstructed using the two-photon decay modes of $\eta$ and
$\pi^0$. Therefore the final state is
$\pi^+\pi^-\gamma\gamma\gamma\gamma$.
For each candidate event, we require that two charged tracks are reconstructed in the MDC and that the polar angles of the tracks satisfy $\left |\cos\theta\right |<0.93$. The tracks are required to pass the interaction point within $\pm10$~cm along the beam direction and within 1~cm in the plane perpendicular to the beam.
All charged particles are assumed to be pions in the analysis. Photon candidates are required to have deposited an energy larger than 25~MeV in the barrel region of the EMC ($\left |\cos\theta\right |<0.8$) and larger than 50~MeV in the endcap region ($0.86<\left |\cos\theta\right |<0.92$). In order to eliminate clusters associated with charged tracks, the angle between the direction of any charged track and a photon candidate must be larger than $10^\circ$.
A requirement on the EMC cluster timing with respect to the event start time ($0 \leqslant T \leqslant 700$ ns) is used to suppress electronic noise and energy deposits unrelated to the event.
Each candidate event is required to have two charged tracks whose net charge is zero and at least four photon candidates that satisfy above criteria.

A four-constraint (4C) kinematic fit, which enforces energy-momentum conservation, is applied assuming the $\pi^+\pi^-\gamma\gamma\gamma\gamma$ hypothesis.
If the number of the selected photons is larger than four, the fit is repeated for all combinations of the photons and the one with the least $\chi^2_{\pi^+\pi^-\gamma\gamma\gamma\gamma}$ value is kept for the further analysis.
Then, a six-constraint (6C) kinematic fit is performed with invariant masses of photon pair combinations constrained to the $\pi^0$ and $\eta$ mass, respectively.
The combination with the smallest $\chi^2_{\pi^+\pi^-\pi^0\eta}$ is selected and the event is retained if $\chi^2_{\pi^+\pi^-\pi^0\eta}<80$.
With this criterion, $97\%$ of all backgrounds can be removed, and the corresponding signal efficiency is $65\%$.

The invariant mass distribution of $\pi^+\pi^-\pi^0\gamma_{\rm low}$,
where $\gamma_{\rm low}$ is the low-energy photon from the pair constrained to the $\eta$ mass, has a peak at the $\eta^\prime$ mass.
The peak is due to background from $J/\psi\rightarrow\gamma\eta^\prime (\eta^\prime\rightarrow\gamma\omega, \omega\rightarrow\pi^+\pi^-\pi^0)$.
The background is removed by requiring $\left |M_{\pi^{+}\pi^{-}\pi^{0}\gamma_{\rm low}}-m_{\eta'}\right |>0.04$~GeV/$c^{2}$,
where $m_{\eta^\prime}$ is the nominal $\eta^\prime$ mass~\cite{Patrignani:2016xqp}.

\begin{figure}[!htbp]
\begin{center}
\includegraphics[width=9.0cm,height=6.0cm]{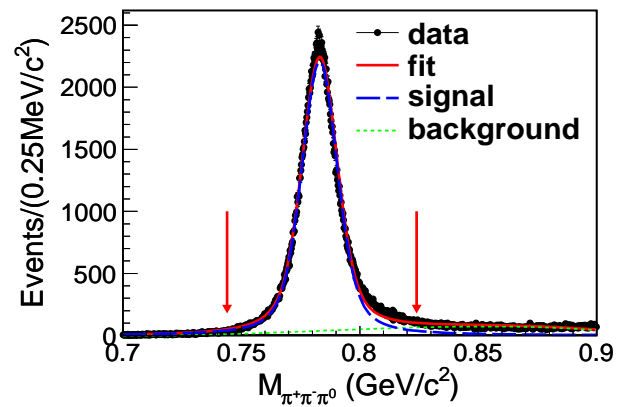}
\caption{Distribution of the $\pi^{+}\pi^{-}\pi^{0}$ invariant mass. The vertical arrows shows the signal region.}\label{fig:4}
\end{center}
\end{figure}

A clear $\omega$ peak is seen in the ${\pi^+\pi^-\pi^0}$ invariant mass distribution after the above requirements, as shown in Fig.~\ref{fig:4}.
The background distribution is smooth and the contribution is low.
The $\omega$ signal region is defined as $\left|M_{\pi^{+}\pi^{-}\pi^{0}}-m_{\omega}\right |<0.04$~GeV/$c^{2}$.
In total, 260,520 candidate events are selected for the $\omega \rightarrow \pi^{+}\pi^{-}\pi^{0}$ Dalitz plot analysis.
Background modes containing a real $\omega$ will not affect the Dalitz plot analysis.
The analysis of the $J/\psi$ inclusive MC sample, using the same selection criteria, shows that the main contributions of peaking background come from $J/\psi\rightarrow\gamma\eta^{\prime}$, $J/\psi\rightarrow\omega\pi^{0}\pi^{0}$ and $J/\psi\rightarrow\omega\pi^{0}$,
which are also $\omega$ processes and only 0.4\% in total,
and therefore can be neglected.
For the non-peaking background, the dominant contribution is from $J/\psi\rightarrow\rho\eta\pi$.
A fit to the $M_{\pi^+\pi^-\pi^0}$ distribution with a Breit-Wigner function convolved with a Gaussian resolution function and added with a second-order polynomial to describe the background, as shown in Fig.~\ref{fig:4}, leads to an estimate of about 4\% of all candidate events to be non-peaking background.

\begin{figure}[!htbp]
\begin{center}
\includegraphics[width=8.0cm,height=6.0cm]{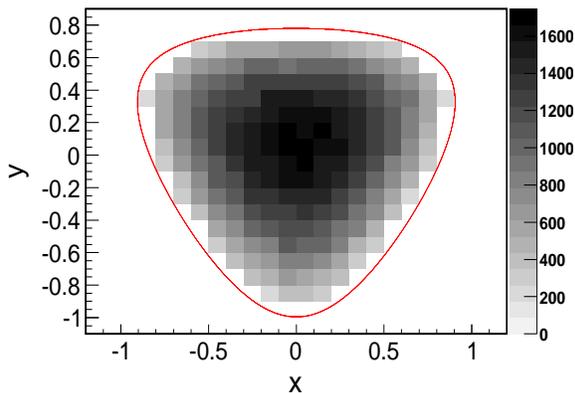}
\caption{Binned Dalitz plot for data expressed using the dimensionless $x$ and $y$
variables. The bins at the Dalitz plot boundary are excluded from
the analysis.}\label{fig:13}
\end{center}
\end{figure}

The Dalitz plot for data and the kinematic boundary (corresponding to $M_{\pi^{+}\pi^{-}\pi^{0}}=m_{\omega}+0.04$~GeV/$c^{2}$)
are shown in Fig.~\ref{fig:13} in terms of the variables $x$ and $y$.
Due to the limited statistics around the kinematic boundary, the Dalitz plot is divided into bins with width $0.1\times 0.1$ in $x$ and $y$,
and then the events in the bins overlapping with the kinematic boundary are not used in the analysis.

\section{Dalitz Plot Analysis}\label{sec:DPA}

\begin{figure}[!htbp]
\begin{center}
\includegraphics[width=8.0cm,height=6.0cm]{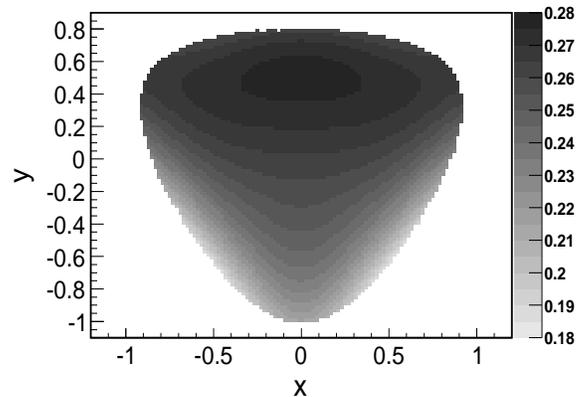}
\caption{Detection efficiency as a function of the two Dalitz plot variables
$x$ and $y$.}\label{fig:hx_rec_fun}
\end{center}
\end{figure}
\begin{figure}[!htbp]
\begin{center}
\includegraphics[width=9.0cm,height=6.0cm]{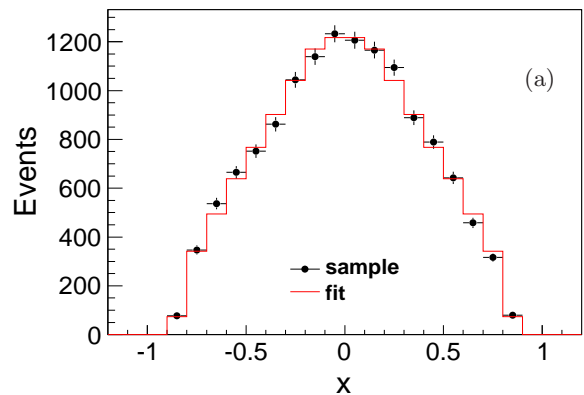}\put (-50,125){(a)} \\
\includegraphics[width=9.0cm,height=6.0cm]{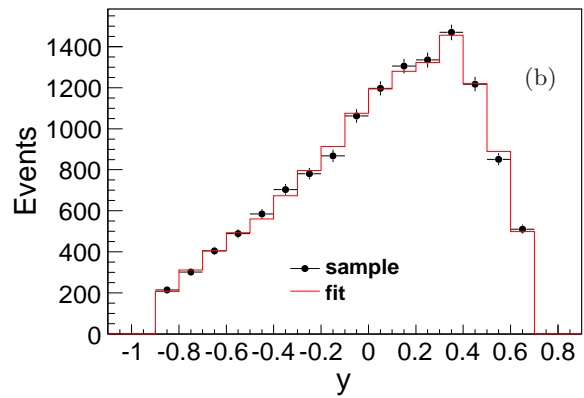}\put (-50,125){(b)}
\caption{Background distribution in the variables
(a) $x$ and (b) $y$ (solid dots). Histograms show the background parametrization used in the Dalitz plot analysis.}\label{fig:hx_bkg_fun}
\end{center}
\end{figure}

We use an unbinned maximum likelihood fit to perform the Dalitz plot analysis, which means a minimization of the logarithmic likelihood function,
\begin{equation}
{- \ln \cal{L}}=-\sum_{i=1}^{N}\ln p(x_{i},y_{i}).\label{eq:LL}
\end{equation}
The probability density function is
\begin{equation}
p(x_{i},y_{i})=(1-f_{B})N_{S}\left |{\cal{M}}\right |^{2}\varepsilon(x,y)+f_{B}N_{B}{\cal{B}}(x,y),\label{eq:probability}
\end{equation}
where $\varepsilon(x,y)$ and ${\cal{B}}(x,y)$ are functions representing the shape of the efficiency and background over the Dalitz plot, respectively;
$N_{B}$ and $N_{S}$ are normalization factors for the background and
signal PDF, respectively, obtained from Monte Carlo integration. The  matrix element squared $\left |{\cal{M}}\right |^{2}$ is defined in Eq.~\eqref{eq:2}. The non-peaking background fraction $f_{B}$ is fixed to 4\%, as discussed earlier.

In order to determine $\varepsilon(x,y)$, a MC sample of 24 million $J/\psi\rightarrow\omega\eta$ events were generated with constant matrix element.
The resulting Dalitz plot distribution for the reconstructed events is shown in Fig.~\ref{fig:hx_rec_fun}.
The resulting efficiency $\varepsilon(x,y)$ is parametrized as a two dimensional polynomial in the variables $x$ and $y$ with maximum degree of terms limited to five
and excluding terms with odd powers of $x$ because of charge symmetry.

Based on the above background study we ignore the peaking
background. For the smooth background under the $\omega$ peak, the
dominant contribution from $J/\psi \rightarrow \rho\eta\pi$ is
studied using a sample of 50 million $J/\psi \rightarrow\rho\eta\pi$ PHSP-generated events.
The $x$ and $y$ projections for the events remaining after the selection are shown in Fig.~\ref{fig:hx_bkg_fun} together with the parametrization used in the analysis,
which is extracted by a fit as ${\cal{B}}(x,y)$ in Eq.~\eqref{eq:probability}.

\section{Fit Results}\label{sec:results}

\begin{figure*}[!htbp]
\begin{center}
\includegraphics[width=9.0cm,height=9.0cm]{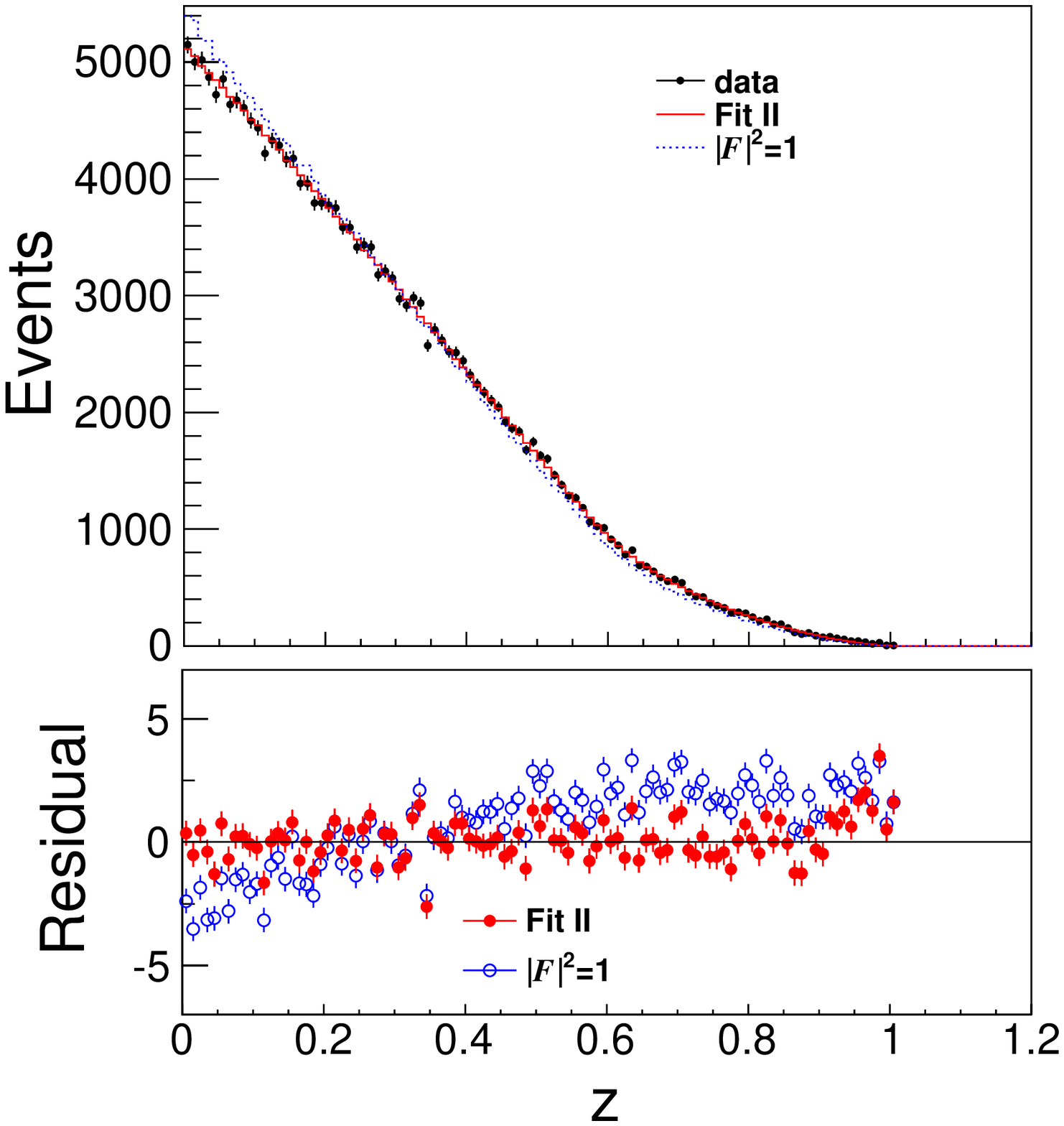}\put (-50,230){(a)}
\includegraphics[width=9.0cm,height=9.0cm]{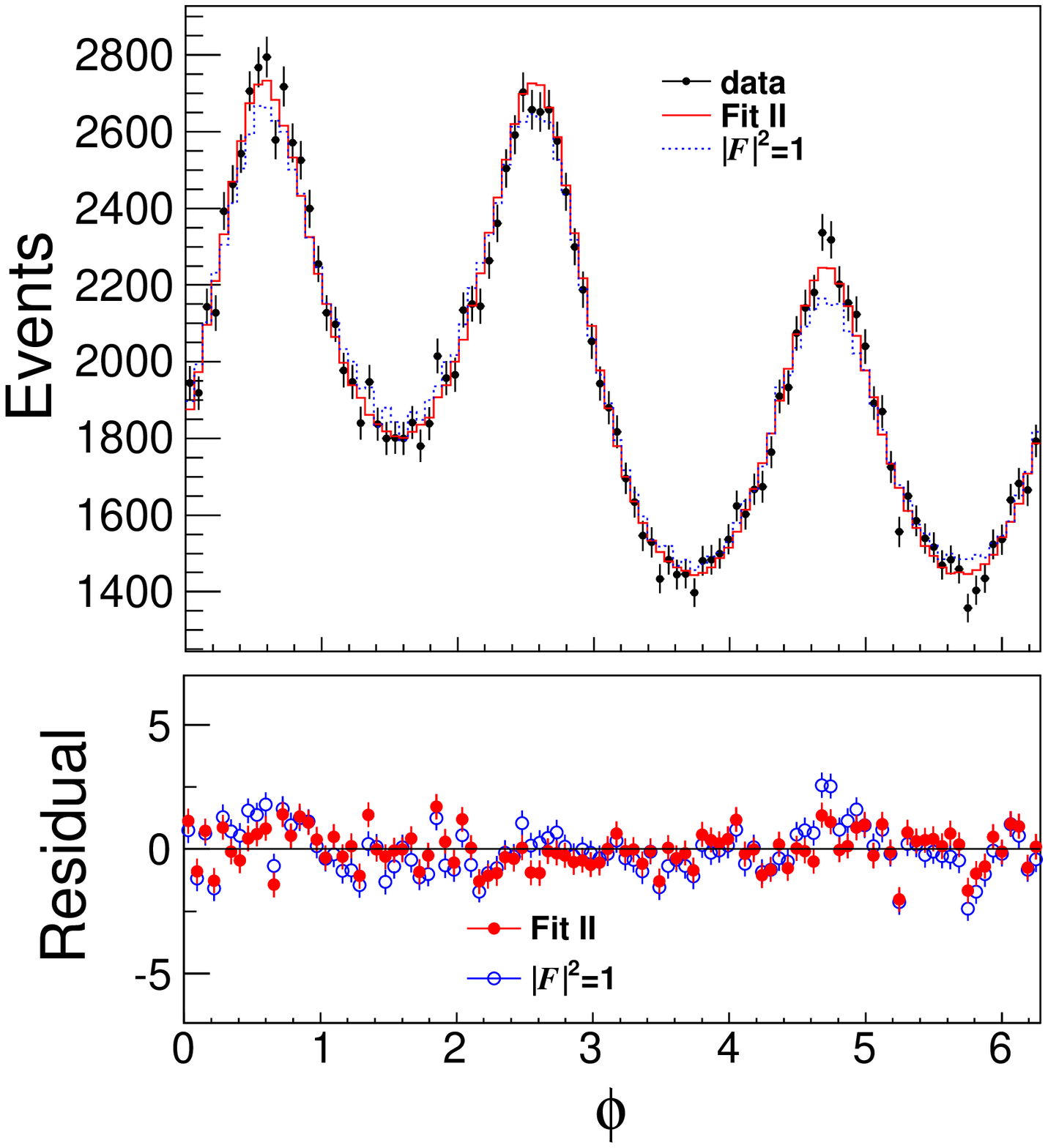}\put (-50,230){(b)} \\
\includegraphics[width=9.0cm,height=9.0cm]{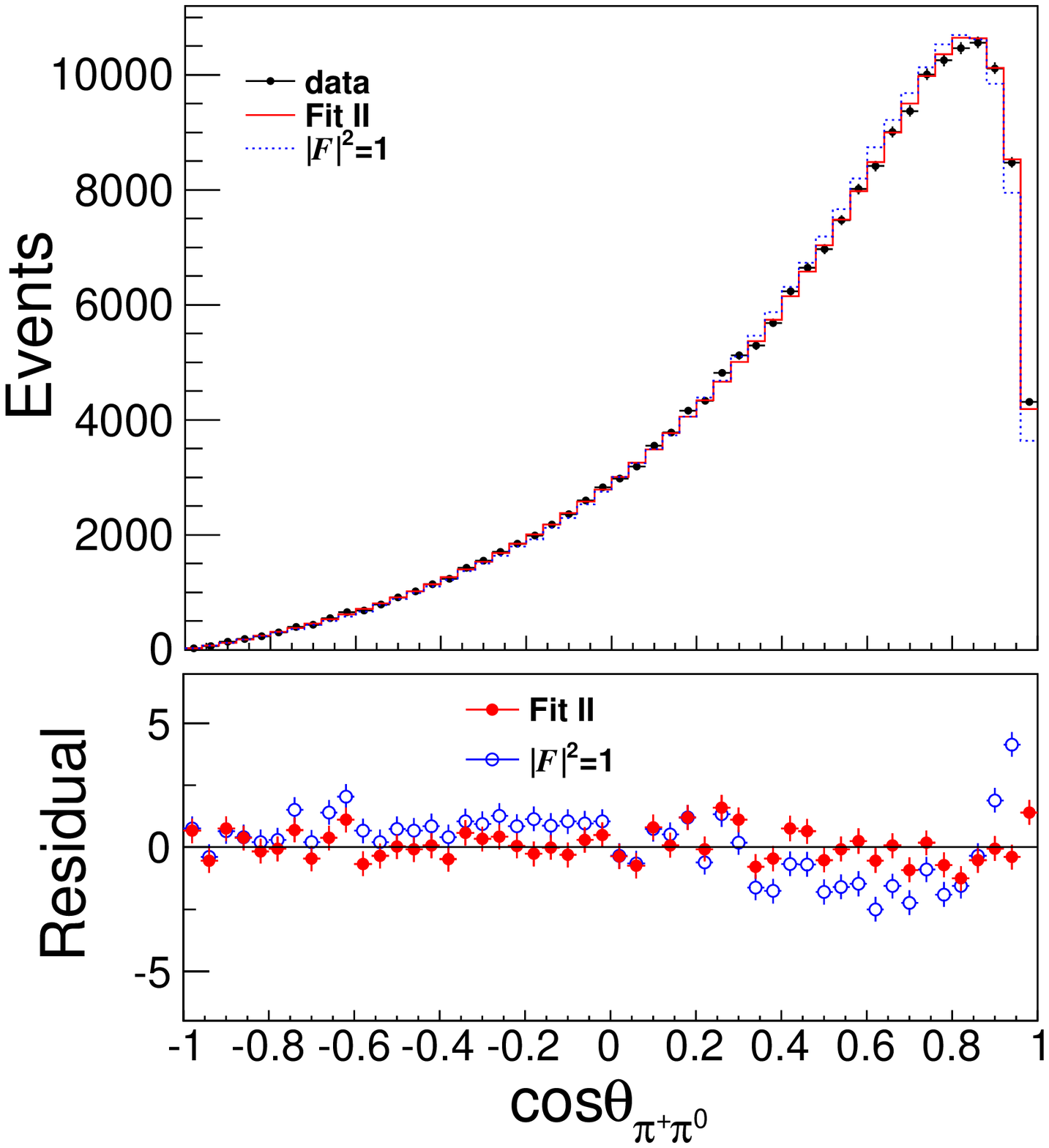}\put (-50,135){(c)}
\includegraphics[width=9.0cm,height=9.0cm]{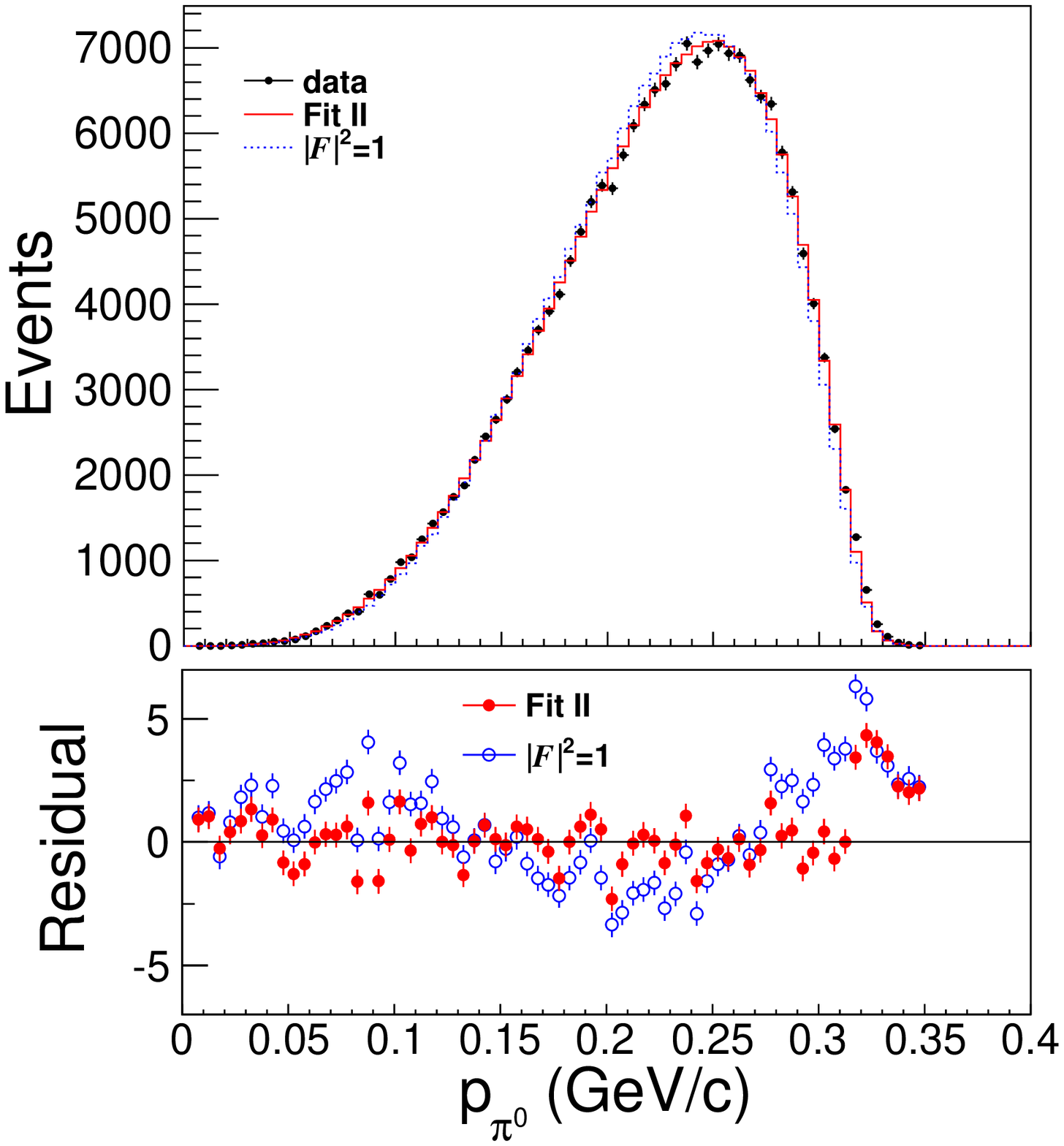}\put (-50,230){(d)}
\caption[]{Data compared to the {\bf{Fit II}} and the amplitude with $|{\cal F}|^2=1$
for different distributions in the $\omega$ rest frame: (a) $z$, (b) $\phi$, (c) $\cos\theta_{\pi^+\pi^0}$, (d) $p_{\pi^-}$, where the black dots with error bars in upper panels are for data, the solid line histograms are for the fit, the dashed line histograms are for $|{\cal F}|^2=1$. The solid and hollow dots in the lower panels denote the residuals for the fit and the $|{\cal F}|^2=1$ assumption, respectively. }
\label{fig:14}
\end{center}
\end{figure*}

With the expected Dalitz plot density in Eq.~(\ref{eq:2}), the fits to the Dalitz plot are performed for different forms of $|{\cal{F}}|^2$.
For the simplest case of $|{\cal{F}}|^2=1$, the discrepancies between the fitted projections and data, in particular for $z$, shown in Fig.~\ref{fig:14}, indicate that the fit is not able to describe data well.

The ansatz $|{\cal{F}}|^2=1+2\alpha z$ ({\bf Fit I})
 results in a much better fit, and $\alpha$ is obtained to be $(132.1\pm6.7)\times10^{-3}$ with a statistical significance of $18.9\sigma$.
The statistical significance is determined by the change of the log-likelihood value and the number of degrees of freedom ($ndf$) in the fit compared to the assumption $|{\cal{F}}|^2=1$.

After including the term $z^{3/2}\sin 3\phi$ ({\bf Fit II}), the fit yields $\alpha=(120.2\pm7.1)\times10^{-3}$ and $\beta=(29.5\pm8.0)\times10^{-3}$. The fit quality is improved, but not significantly, as implied by the statistical significance of the coefficient $\beta$, $4.3\sigma$,
which is calculated by comparing to the likelihood and $ndf$ for {\bf Fit I} (the difference between $ndf$ of {\bf Fit I} and {\bf Fit II} is 1).
A comparison (Fig.~\ref{fig:14}) between fit and data for the projections in different variables and the residuals shows that the fit can provide a good description of data.
For the two-dimensional distribution in $(x,y)$, the $\chi^{2}/ndf$ is $832/805$.

An alternative fit ({\bf Fit III}), introducing the term proportional to $z^{2}$, is also performed.
It turns out that the fitted values of $\alpha$ and $\beta$ summarized in Table~\ref{tab:4} are in agreement with those of {\bf Fit II},
while the coefficient $\zeta$, $(22\pm29){\times}10^{-3}$, is consistent with zero and the corresponding statistical significance is only $1.3\sigma$.
Therefore it is justified to ignore the higher order contributions based on the present statistics.

\begin{table}[!htbp]
\begin{center}
\caption{Summary of the fit results for the different forms of $|{\mathcal{F}}|^2$.}\label{tab:4}
\begin{small}
\begin{tabular}{lccc}\hline\hline
   & $\alpha \times 10^{3}$ & $\beta \times 10^{3}$ & $\zeta \times 10^{3}$ \\\hline

  \bf{Fit I} & $132.1\pm6.7$ & $ $ & $ $ \\

  \hline

  \bf{Fit II} & $120.2\pm7.1$ & $29.5\pm8.0$ & $ $ \\

  \hline

  \bf{Fit III} & $111\pm18$ & $25\pm10$ & $22\pm29$ \\

  \hline\hline
\end{tabular}
\end{small}
\end{center}
\end{table}

\section{Systematic Uncertainties}\label{sec:systematic_uncertainties}

To ensure the stability and reliability of the results, input-and-output checks with toy MC samples are used. The systematic shifts between input and output parameters are taken as the uncertainties related to the fit method. For the uncertainty due to the efficiency parametrization, we perform alternative fits by using, instead of the polynomial efficiency function, the average efficiencies in the Dalitz plot bins. The changes of the results with respect to the standard analysis are treated as the systematic uncertainty.

The impact of the resolution in $x$ and $y$ is studied by implementing the resolution functions, numerically convolving them with the probability density function without considering the correlation between $x$ and $y$. Since the resolution value is much smaller than the bin widths of $x$ or $y$, we find that the results almost do not change by including the resolutions. Thus the systematic uncertainty from this source is neglected.

Differences between the data and MC samples for the tracking efficiency of charged pions are investigated using $J/\psi \rightarrow p\bar{p}\pi^{+}\pi^{-}$ decays.
A momentum-$\cos\theta$-dependent two-dimensional correction is obtained for the charged pions in the MC events.
Similarly, a momentum-dependent correction for the $\pi^{0}$ efficiency in the MC sample is obtained from $J/\psi \rightarrow \pi^{+}\pi^{-}\pi^{0}$ decays.
The fits to extract the parameter values are repeated, taking into account the efficiency correction for charged and neutral pions.
The change of the results with respect to the default fit result is assigned as a systematic uncertainty.

The systematic uncertainty for the signal region ($\left|M_{\pi^{+}\pi^{-}\pi^{0}}-m_{\omega}\right|<0.04$~GeV/$c^2$) is estimated by replacing the nominal selection with a very loose requirement ($\left|M_{\pi^{+}\pi^{-}\pi^{0}}-m_{\omega}\right|<0.1$~GeV/$c^2$).  The systematic uncertainty due to the $\eta^\prime$ veto is evaluated by excluding this requirement, but including this contribution estimated from the MC simulation.

To evaluate the uncertainty associated with the 6C kinematic fit, the approach described in detail in Ref.~\cite{Ablikim2013} is used to correct the track helix parameters of the MC simulation to improve agreement between data and MC simulation. In this analysis, we find this correction to have some impact on the results. Therefore, we take the result with correction as the nominal one, and the difference between the result with and without correction as the systematic uncertainty from the kinematic fit.

As we mentioned above, the background events under the $\omega$ peak are estimated with the MC events of $J/\psi\rightarrow\rho\eta\pi$. To estimate the impact from the background uncertainty, we perform an alternative fit by determining the background from the $\omega$ mass sidebands of the experimental $M_{\pi^{+}\pi^{-}\pi^{0}}$ distribution ($0.08$ GeV/$c^{2}<\left|M_{\pi^{+}\pi^{-}\pi^{0}}-m_{\omega}\right|<0.12$~GeV/$c^2$). The differences between the results for the extracted parameters to the nominal ones are taken as the systematic uncertainties.

The systematic uncertainty from the above sources for $\alpha$ in {\bf{Fit I}} is $4.1\times10^{-3}$.
And for {\bf{Fit II}}, the systematic uncertainties are summarized in Table~\ref{tab:12}
($\sigma_{\alpha}$ and $\sigma_{\beta}$ denote absolute uncertainties of $\alpha$ and $\beta$, respectively).
The total systematic uncertainty is determined by adding all contributions in quadrature.

\begin{table}[!htbp]
\begin{center}
\caption{Systematic sources and their contributions for {\bf{Fit II}}.}\label{tab:12}
\begin{small}
\begin{tabular}{lcc}\hline\hline
  Source & $\sigma_{\alpha}\times10^{3}$ & $\sigma_{\beta}\times10^{3}$ \\\hline
  Fit  bias & $1.8$ & $1.3$ \\
  Efficiency parametrization & $0.4$ & $1.6$ \\
  Charged track reconstruction & $1.3$ & $1.8$ \\
  $\pi^{0}$ reconstruction & $1.2$ & $1.1$ \\
  ${\omega}$ signal region & $2.3$ & $3.0$ \\
  ${\eta'}$ veto & $0.9$ & $2.5$ \\
  Kinematic fit & $0.9$ & $0.1$ \\
  Background & $0.8$ & $2.1$ \\\hline
  Total  & $3.8$ & $5.3$ \\
  \hline\hline
\end{tabular}
\end{small}
\end{center}
\end{table}

\section{Summary}\label{sec:summary}

\begin{table*}[!htbp]
\begin{center}
\caption{Predictions and fit results for the ${\cal{F}}$ parametrizations. The predictions are from
Danilkin {\it et al.} \cite{Danilkin:2014cra}, Niecknig {\it et al.}~\cite{Niecknig:2012sj}, and Terschl{\"u}sen {\it et al.}~\cite{Terschlusen:2013iqa}.
Theoretical predictions without incorporating crossed-channel effects are indicated by \underline{\bf{w/o}}
and those with crossed-channel effects by \underline{\bf{w}}.}\label{tab:19}
\begin{tabular}{|c|c|cc|cc|c|c|}
  \hline\hline
  \multirow{3}{*}{ } & Para. & \multicolumn{5}{c|}{Theoretical Predictions} & \multicolumn{1}{c|}{Experiment} \\\cline{3-8}
                        & $\times$ &\multicolumn{2}{c|}{Ref.~\cite{Danilkin:2014cra}}
&\multicolumn{2}{c|}{Ref.~\cite{Niecknig:2012sj}}& Ref.~\cite{Terschlusen:2013iqa}& \multirow{2}{*}{BESIII} \\
                        & $10^{3}$ & {\bf{w/o}} &  {\bf{w}} & {\bf{w/o}} & {\bf{w}} &  & \\
  \hline\hline
  \multirow{1}{*}{\bf{Fit I}} & $\alpha$ & $136$ & $94$ & $(137,148)$ & $(84,96)$ & $202$ & $132.1 \pm 6.7 \pm 4.6$ \\
  \hline
  \multirow{2}{*}{\bf{Fit II}} & $\alpha$ & $125$ & $84$ & $(125,135)$ & $(74,84)$ & $190$ & $120.2 \pm 7.1 \pm 3.8$ \\
                     & $\beta$ & $30$ & $28$ & $(29,33)$ & $(24,28)$ & $54$ & $29.5 \pm 8.0 \pm 5.3$ \\
  \hline\hline
\end{tabular}
\end{center}
\end{table*}

Using a sample of 1.3 billion $J/\psi$ events collected with the BESIII detector, we perform a Dalitz plot analysis of the decay $\omega\rightarrow\pi^+\pi^-\pi^0$ using $J/\psi\rightarrow\omega\eta$ decays. The comparison to the theoretical predictions for different sets of the fitted parameters  is given in Table~\ref{tab:19}.
The predictions are from dispersive analyses by Niecknig {\it et al.}~\cite{Niecknig:2012sj} and by Danilkin {\it et al.}~\cite{Danilkin:2014cra}.
Both analyses give predictions for two cases: without incorporation of crossed-channel effects (1)
and with incorporation of crossed-channel effects (2).
In addition, predictions from a Lagrangian based study with pion-pion rescattering effects by Terschl{\"u}sen {\it et al.}~\cite{Terschlusen:2013iqa} are shown.
The parameters determined experimentally agree with those predicted within the dispersion framework. Our data clearly
show that the Dalitz plot distribution deviates from the $P$-wave phase space (\emph{i.e.,} $\left|{\cal{F}}\right|^{2}=1$).
The value of the parameter $\alpha$, $\alpha=(134.9\pm6.8\pm4.1)\times10^{-3}$, is established with very good precision and consistent with the dispersive calculations, $\alpha=136\times10^{-3}$.
Further introduction of the parameter $\beta$ improves the significance only a little.
With present statistics, other higher-order parameters are not necessary to describe the data of $\omega\rightarrow\pi^{+}\pi^{-}\pi^{0}$
since the parameter value from the fit, \emph{e.g.} $\zeta$, is consistent with zero.
The fitted parameter values are consistent with the theoretical predictions without incorporating crossed-channel effects.
This may indicate that the contribution of the crossed-channel effects is overestimated in the dispersive calculations.

\begin{acknowledgments}
The BESIII collaboration thanks the staff of BEPCII and the IHEP computing center for their strong support.
This work is supported in part by National Key Basic Research Program of China under Contract No. 2015CB856700;
National Natural Science Foundation of China (NSFC) under Contracts Nos. 11335008, 11425524, 11625523, 11635010, 11675184, 11735014;
the Chinese Academy of Sciences (CAS) Large-Scale Scientific Facility Program;
the CAS Center for Excellence in Particle Physics (CCEPP);
Joint Large-Scale Scientific Facility Funds of the NSFC and CAS under Contracts Nos. U1532257, U1532258, U1732263;
CAS Key Research Program of Frontier Sciences under Contracts Nos. QYZDJ-SSW-SLH003, QYZDJ-SSW-SLH040;
100 Talents Program of CAS;
INPAC and Shanghai Key Laboratory for Particle Physics and Cosmology;
German Research Foundation DFG under Contracts Nos. Collaborative Research Center CRC 1044, FOR 2359;
Istituto Nazionale di Fisica Nucleare, Italy;
Koninklijke Nederlandse Akademie van Wetenschappen (KNAW) under Contract No. 530-4CDP03;
Ministry of Development of Turkey under Contract No. DPT2006K-120470;
National Science and Technology fund;
The Swedish Research Council;
U. S. Department of Energy under Contracts Nos. DE-FG02-05ER41374, DE-SC-0010118, DE-SC-0010504, DE-SC-0012069;
University of Groningen (RuG) and the Helmholtzzentrum f{\"u}r Schwerionenforschung GmbH (GSI), Darmstadt.
\end{acknowledgments}

\bibliographystyle{apsrev4-1}
\bibliography{draft_omega23pi}

\end{document}